\documentclass[twocolumn, superscriptaddress]{revtex4-1}
\pdfoutput=1

\usepackage[utf8]{inputenc}
\usepackage{amsmath}
\usepackage{graphicx}
\usepackage{soul} % highlighting

\DeclareMathOperator{\sech}{sech}

\newcommand{\delzeta}{\partial_{\zeta}}

\newcommand{\delttau}{\partial^{2}_{\tau \tau}}
\newcommand{\deltttau}{\partial^{3}_{\tau \tau \tau}}
\newcommand{\abs}[1]{\left| #1 \right|}

\begin{document}

% Title page
\title{Emulation of Fabry-Perot and Bragg resonators with temporal optical solitons.}

\author{T. Voytova}
\email{Corresponding author: t.voytova@phoi.ifmo.ru}
\affiliation{ITMO University 197101, Kronversky pr. 49, St. Petersburg, Russian Federation}

\author{I. Oreshnikov}
\affiliation{ITMO University 197101, Kronversky pr. 49, St. Petersburg, Russian Federation}

\author{A.V. Yulin}
\affiliation{ITMO University 197101, Kronversky pr. 49, St. Petersburg, Russian Federation}

\author{R. Driben}
\affiliation{ITMO University 197101, Kronversky pr. 49, St. Petersburg, Russian Federation}
\affiliation{Department of Physics and CeOPP, University of Paderborn, Warburger St. 100, D-33098 Paderborn, Germany}

\date{\today}

% Abstract
\begin{abstract}
The scattering of weak dispersive waves (DW) on several equally spaced temporal solitons is studied. It is shown by systematic numerical simulations that the reflection of the DWs from the soliton trains strongly depends on the distance between the solitons. The dependence of the reflection and transmission coefficients on the inter-soliton distance and the frequency of the incident waves is studied in detail, revealing fascinating quasi-periodic behavior. The analogy between the observed nonlinear phenomena in temporal domain and usual Fabry-Perot and Bragg resonators is discussed.
\end{abstract}
\maketitle

\section{Introduction}

Propagation of temporal solitons in optical fibers is very popular
research area due to its rich fundamental physics and variety of
possible applications ranging from telecommunications to
supercontunuum light sources \cite{book}. When solitons are launched
in the vicinity of zero dispersion point \cite{Wai} radiation of
resonant (Cherenkov) dispersive waves (DWs) takes place
\cite{Akhmediev}. These DWs play important role in the dynamics of the emitting soliton via spectral recoil \cite{Recoil} and they also affect the neighboring pulses \cite{DribenMitschke}. Very recently resonant radiation of oscillating second order solitons was investigated \cite{DribenYulin} revealing interesting structure of radiation band consisting of multiple frequency peaks. Similar radiation features were reported with optical pulses propagating in fibers with alternating dispersion \cite{KudlinskiA, KudlinskiB}.

A special effort was devoted to treating interaction of soliton with external DW incident on it. Scattering of weak DWs on fundamental
solitons was reported in numerous publications \cite{yulin, pre, Efimov, Efimov2, Skryabinoverview, Conforti}. Also scattering of external dispersive waves on high order soliton was considered in \cite{2soliton} and dark solitons \cite{darksolitons}. In recent works \cite{Resonator, Resonator2, Kudlinski} it was shown that DWs can be trapped between  solitons and that this DWs can cause efficient inter-soliton interaction when the distance between the solitons is much larger compared to the soliton width.

Interaction of more intense DWs colliding with solitons leads to acceleration or deceleration of solitons with the resulting frequency upshifts or downshifts \cite{DribenMitschke, YulinPRE, Demircan1, Demircan2, Tartara}. The following dynamics suggested prospective applications as all-optical switch \cite{Demircan1, Tartara} or an alternative technique for generation of broad and coherent supercontinuum \cite{Demircan2} without the high order soliton fission. Related phenomena were very recently realized in integrated nanophotonic waveguides \cite{Kuyken}.

% AVY:
% The present work aims to show that solitonic trap can work as resonator very much similar to classical Fabry-Perot resonator. We also show that a train of optical solitons in fibers can work as a Bragg mirror for DWs of low intensity.

% RD:
% The present work aims to exploit further interaction of DWs with the ensembles of solitons and to demonstrate emulation of famous Fabry-Perot resonators and Bragg reflectors by nonlinear optics in temporal domain. There were numerous works in spatial domain nonlinear optics to mimic the functionality of the classical optical devices such as \cite{Spatial1, Spatial2}. However the discussed below effects persist purely in temporal domain optics where the asymmetric third order dispersion (TOD) exists.

%Final: I've skipped references to Spatial1 and Spatial2 since -- dealing with soliton scattering on an external potential -- they are not exactly related to our work. Citing them could mislead a lazy reviewer.
The present work aims to further explore interaction of DWs with the ensembles of solitons and to demostrate emulation of famous Fabry-Perot resonators, as well as Bragg reflectors, in temporal domain. The effects discussed below rely not only on refractive index change due to optical Kerr effect, but also on significant (but small) higher-order dispersion in the fiber.

To describe propagation of light in fibers close to zero-dispersion wavelength we adopt a simplified model of Nonlinear Schrodinger Equation with additional third-order dispersion (TOD) term. For a complex-valued envelope $u(\zeta, \tau)$ it reads \cite{book}:
\begin{equation}
  \label{eq:NLSE}
  i \delzeta u
    + \frac{1}{2} \delttau u
    - i\delta_3 \deltttau u
    + \abs{u}^2 u = 0,
\end{equation}
where $\tau$ is dimensionless time $\tau = t / T_0$ for a characteristic time scale $T_0$, $\zeta = z / L_{D}$ with dispersion length $L_{D} = |{T_{0}}|^2 / \beta_{2}$. The coefficient $\delta_{3}=\beta_{3}/(6T_0\beta_{2})$ represents the relative strength of the TOD. It is known that Raman effect significantly influences propagation of femtosecond pulses. However, for hollow core fibers filled with Raman free gases, or for the case of relatively long pulses the Raman effect can be neglected. Aiming to demonstrate the effect more clearly we report the results only for the case when Raman term is disregarded.

We study the evolution of and initial condition given as a sum of a soliton ensemble $u_{se}$ and a long dispersive wave envelope $u_{dw}$ situated far away from the solitons
\begin{equation}
  \label{eq:InitialCondition}
  u(0, \tau) = u_{se}(0, \tau) + u_{dw}(0, \tau),
\end{equation}
where the soliton ensemble is given by
\begin{equation}
  \label{eq:SolitonResonator}
  u_{se}(0, \tau)
    = u_0 \sech{u_0 (\tau + \Delta/2)}
    + u_0 \sech{u_0 (\tau - \Delta/2)}
\end{equation}
in the case of a soliton cavity and
\begin{equation}
  \label{eq:SolitonChain}
  u_{se} = \sum_{n = 0}^{N} u_0 \sech{u_0 (\tau - n \cdot \Delta)}
\end{equation}
in the case of an $N$-soliton chain. Initial shape of incident dispersive wave is given by a super-Gaussian pulse with a relative frequency detuning of $\omega_{inc}$
\begin{equation}
  \label{eq:IncidentDW}
  u_{dw} = A_{inc}
    \cdot \exp\left(
      -(\tau - \tau_0)^{\gamma} / w^{\gamma}
    \right)
    \cdot \exp(- i \omega_{inc} \tau),
\end{equation}
with $|\tau_{0}| \gg \Delta$

\section{Reflection from a two-soliton cavity}

First we consider how the broad dispersive pulse collide and scatters on a pair of well separated non-interacting solitons.
% To avoid the effect of trajectory change of solitons via their collisions with DWs \cite{DribenMitschke, Demircan1} we launch the soliton pair together with broad DWs of very low intensity.
To study the scattering of the soliton pair we performed a series of numerical runs and measured the total number of photons reflected from the solitons. The intensity of DWs was chosen approximately three order of magnitude lower then the intensity of the solitons in order to avoid the soliton changing their trajectory after interacting with the DWs \cite{DribenMitschke, Demircan1}. The ratio of the number of the reflected photons to the number of photons in the initial pulse gives us the value of the reflection coefficient $R$. The transmission coefficient $T$ is introduced as a ratio of the photons in the waves passed trough the solitons cluster to the total number of photons in the initial pulse. In both of those definitions we assume that the number of the photons is proportional to the integral of optical intensity $\int |u|^2 dt$.

\begin{figure}[h]
  \centering
  \includegraphics[width=8.8cm]{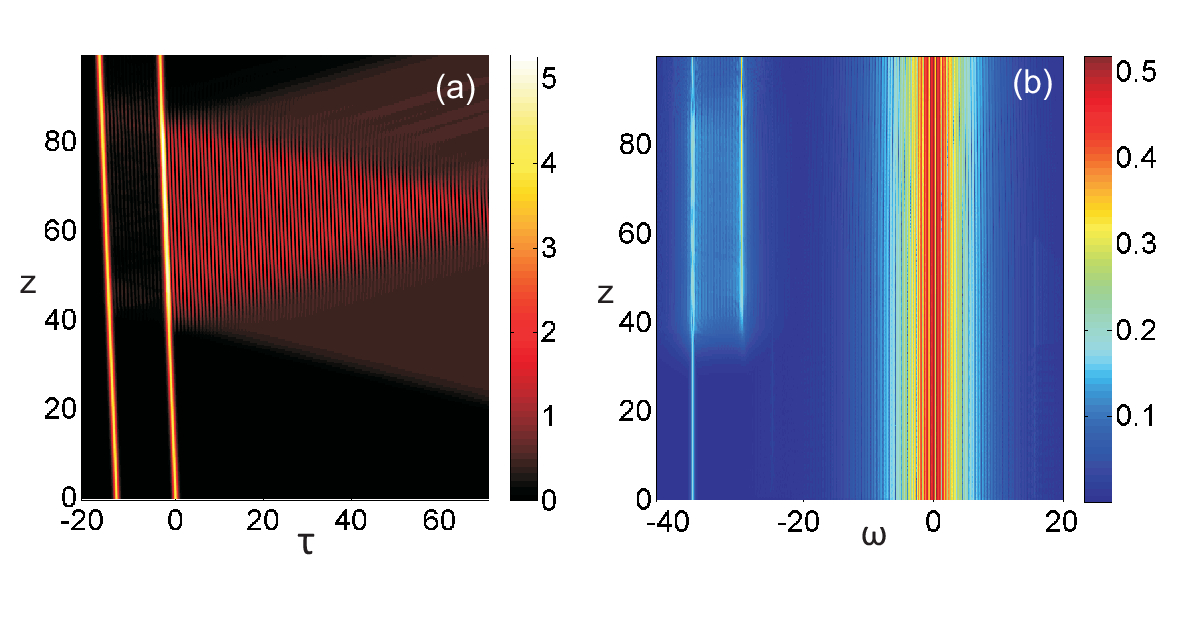}
  \caption{(Color online) The dynamics of the solitons and dispersive pulse collision. (a) The intensity evolution in the (z,$\tau$)-plane. (b) The pulse intensity in the distance-frequency diagram. The parameters are $w=100;~\omega_{inc}=-37; ~A_{inc}=0.007; ~N=2; ~u_0=2; ~\Delta=13.2$.}
  \label{fig:1}
\end{figure}

Typical dynamics of interaction of external weak and broad DW with a pair of solitons is presented in Fig.~\ref{fig:1}. Dispersive wave with duration $w=100$ and an initial time delay $\tau_0=120$ propagates toward the solitons, collides and gets partially reflected from the fist soliton. For the given set of parameters the interaction between the DW and the solitons starts at the distance $z \approx 40$. As it clearly seen from the Fig.~\ref{fig:1}b the reflected part of DW has a frequency $\omega_{ref}\approx-29.4$, as it is predicted by the resonance condition derived in \cite{yulin}. The transmitted part of DW does not experience any frequency shift and continues to propagate with initial detuning $\omega_{inc}=-37$. The wave passed through the first soliton get reflected the second soliton, then this wave of the frequency $\omega_{ref}\approx-29.4$ scatters on the first soliton one more time, producing a reflected wave with the frequency $\omega=-37$. Thus, after a number of reflections, the field between the solitons can be represented as two counter-propagating waves with different frequencies and the same propagation constant. This process is analogous to what happens in conventional Fabry-Perot resonator.

\begin{figure}[h]
  \centering
  \includegraphics[width=8.8cm]{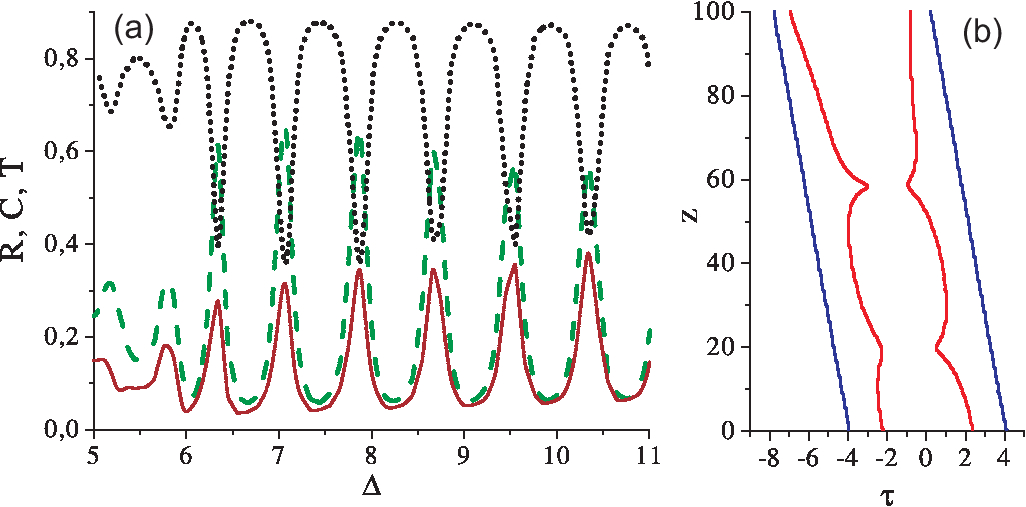}
  \caption{(Color online) (a) The dependence of reflection ($R$, black dotted curve), trapping ($C$, solid red curve) and transmission ($T$, green dashed curve) coefficients on the temporal separation between the solitons for $w=50;~\omega_{inc}=-37; ~A_{inc}=0.007; ~u_0=2; ~N=2$. (b) The solitons trajectories at different initial temporal separation: interacting case with $\Delta=5.0$ corresponds to red curves, noninteracting case with $\Delta=8.0$ corresponds to blue curves.}
  \label{fig:2}
\end{figure}

% Something is off about this paragraph. Maybe it's all the "dependencies".
By direct numerical simulation we studied how scattering of DWs from the soliton resonator depends on the distance between the solitons. The corresponding dependencies of the reflection and the transmission coefficients as function of the distance between the solitons are shown in Fig.~\ref{fig:2}a. It is seen that the dependencies of the reflection and transmission coefficients on the distance between the solitons have oscillatory behaviour, very much similar to the dependencies of the reflection and transmission coefficients on the distance between the interfaces in conventional Fabry-Perot resonators. Representing the field in the form of two counter-propagating waves it is straightforward to obtain the expression for the  period of the oscillations $\Delta \tau = 2\pi/|\omega_{inc}-\omega_{ref}|$. For our parameters the analytical period of the reflection coefficient oscillations is equal to $\Delta \tau \approx 0.83$, which matches well the period of the numerically calculated curve.

We also measured how the maximum number of photons trapped between the solitons depends on the inter-soliton distance (solid line in the Fig.~\ref{fig:2}a). One can see that this behaviour is also periodic and that the maxima of transmission coincides with the maxima of the number of trapped photons. This clearly supports the analogy between the present setting and classical Fabry-Perot resonators.

In case of very short inter-soliton separation $\Delta$ the periodicity is reduced due to the direct interaction between the tails of the two solitons. Fig.~\ref{fig:2}(b) shows the trajectories of solitons with different initial temporal separation: blue curves corresponds to  well separated noninteracting solitons with $\Delta=8$, red curves - to interacting solitons with $\Delta=5$. Further we will consider the interaction of DW with a well separated solitons.

\begin{figure}[h]
  \centering
  \includegraphics[width=8.8cm]{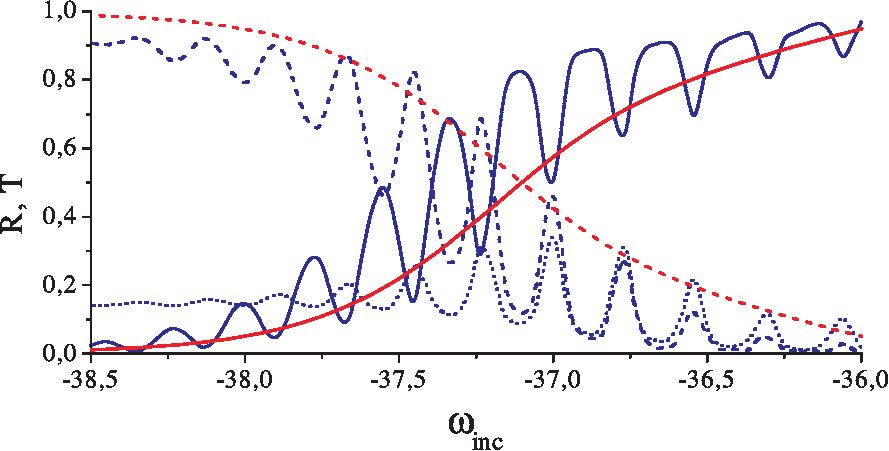}
  \caption{(Color online) The reflection (solid curves), trapping (dotted curve) and transmission (dashed curves) coefficients depending on the incident dispersive wave frequency in the case of a solitary soliton (red curves) and a pair of solitons (blue curves) $w=50;~\Delta=12;~A_{inc}=0.007$}
  \label{fig:3}
\end{figure}

Another way to test the resonant properties of a soliton clusters is to measure the dependence of the reflection and the transmission coefficients on the frequency of the incident probing wave. In Fig.~\ref{fig:3} we compare the reflection and transmission coefficients of a single soliton with the same characteristics of the two-soliton cluster. The curve for the reflection coefficient of a single soliton is calculated numerically but can be approximated well by the formula derived in \cite{Wang}:
\begin{equation}
  \label{eq:ApproxReflectionCoefficient}
  R(\omega_{inc})= \frac{
    \cosh^2 \left(
      \frac{\sqrt{15}}{2} \pi
    \right)
  }{
    \sinh^2 \left(
      \pi(\omega_{inc}-\omega_0)t_0
    \right) +
    \cosh^2 \left(
      \frac{\sqrt{15}}{2} \pi
    \right)
  },
\end{equation}
where $t_0 = 1 / \sqrt{2q}$ is proportional to soliton duration and $\omega_0$ is the zero group velocity dispersion point. One can see that the reflection-transmission coefficients for a single soliton and for a two-soliton clusters behaves differently. In the case of the soliton cluster the reflection coefficient shown pronounced oscillations imposed on a smooth background resembling the dependence of the reflection coefficient of a single soliton on the frequency.

\section{Reflection from a soliton train}

It was shown above that the reflection of DWs from a soliton cluster can demonstrate a resonant nature and thus a soliton pair can be seen as a Fabry-Perot resonator for dispersive waves. Thinking in a similar way, we can expect that a soliton train can work as a Bragg mirror for the incident dispersive waves. This effect allows one to obtain strong and frequency selective reflection from soliton trains in the range of parameters when the reflection from a single soliton is weak. So, strong reflection can be obtained even in the case of low intensity solitons and the efficiency of the reflection can be controlled by the inter-soliton distance of the soliton trains. This effect can possibly be used for the spectroscopy of the soliton trains, when information about the parameters of the soliton trains is extracted from the scattering data of the probing DW.

\begin{figure}[h]
  \centering
  \includegraphics[width=8.8cm]{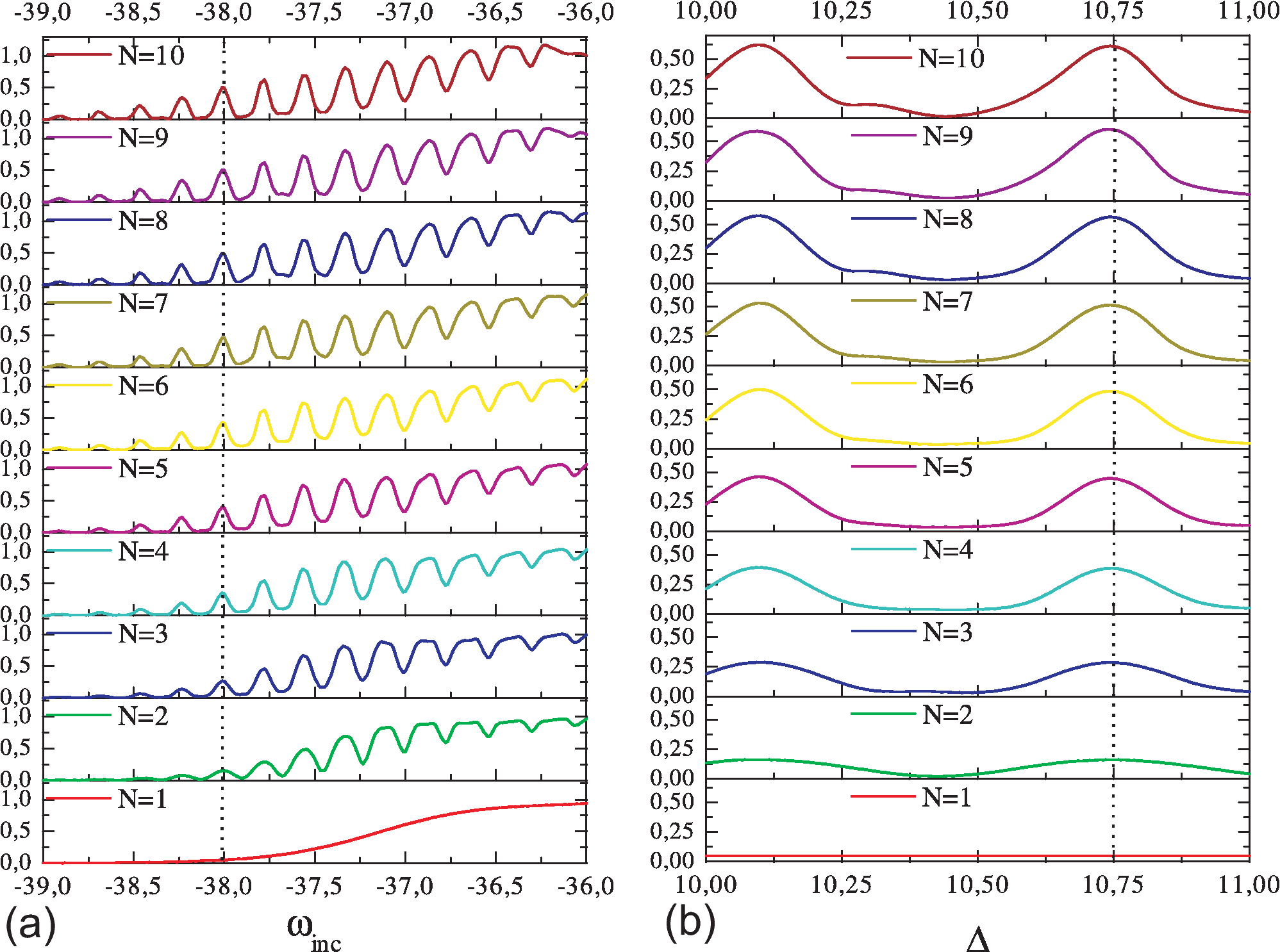}
  \caption{The dependency of the reflection coefficient of the probe pulse from the lattices consisting of different numbers of solitons $w=50; A_{inc}=0.007$. (a) $\Delta=10.74$. (b) $\omega_{inc}=-38$.}
  \label{fig:4}
\end{figure}

The numerically calculated dependencies of the reflection coefficients on the frequency of the incident waves are shown in Fig.~\ref{fig:4}a for the soliton trains consisting of different number of solitons. It is seen that the reflection from a single soliton is negligible for the frequencies lower than the one marked by a vertical dashed line. However, increasing the number of solitons, we can enhance the reflection greatly. For example, for the frequency $\omega_{inc}=-38$ the reflection from a single soliton is more than $n=10$ times weaker comparing to the reflection from a soliton train consisting of $N=10$ solitons.

\section{Conclusions}

To conclude, we briefly summarize the main results reported in the Letter. It was shown that in the presence of the high order dispersion clusters of well separated intense optical pulses (solitons) can work as dynamical Fabry-Perot resonators. The dependence of the reflection and transmission coefficients on the inter-soliton distance and on the frequency of the incident waves is studied in details and the analogy between the soliton clusters and classical Fabry-Perot resonators is demonstrated. The case of multi-soliton trains is also discussed. It is found that Bragg resonance can appear in the soliton trains. This resonance can greatly enhance the reflection from solitons trains in the case when the reflection from an individual soliton is extremely weak. The reported effects are frequency selective, and thus can be used for the spectroscopy of the soliton trains by weak DWs.

\section{Funding information}
RD and AVY were financially supported by the Government of the Russian Federation (Grant 074-U01) through ITMO Early Career Fellowship.

\clearpage

\end{document}